\documentclass{aa}  
\usepackage{graphicx}
\usepackage{txfonts}
\usepackage{bm}
\usepackage{subcaption}
\usepackage{xcolor}

\usepackage{placeins}

\newcommand{\rmm}[1]{\xspace} 

\newcommand{\idefix}{\textsc{Idefix}}
\newcommand{\athena}{\textit{Athena++}}

\newcommand{\orcidlink}[1]{\protect\href{https://orcid.org/#1}{\protect\includegraphics[width=8pt]{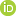}}}

\usepackage{natbib,twoopt}
\usepackage[breaklinks=true]{hyperref} 

\begin{document}

   \title{Spiral-wave-driven accretion in quiescent dwarf nov\ae}

   \author{M. Van den Bossche\orcidlink{0000-0003-4755-9875}
          \inst{1}
          \and
          G. Lesur\orcidlink{0000-0002-8896-9435}\inst{1}
          \and
          G. Dubus\orcidlink{0000-0002-5130-2514}\inst{1}
          }

   \institute{Univ. Grenoble Alpes, CNRS, IPAG, 38000 Grenoble, France\\
              \email{marc.vanden-bossche@univ-grenoble-alpes.fr}
                        }

   \date{Received XXX; accepted April 26, 2023; in original form \today}

  \abstract
  {In dwarf nov\ae\ (DNe) and low-mass X-ray binaries, the tidal potential excites spiral waves in the accretion disc. Spiral-wave-driven accretion may be important in quiescent discs, where the angular momentum transport mechanism has yet to be identified. Previous studies were limited to unrealistically high temperatures for numerical studies or to specific regimes for analytical studies. }
   {We perform the first numerical simulation of spiral-wave-driven accretion in the cold temperature regime appropriate to quiescent discs, which have Mach numbers $\gtrsim 100$.}
   {We used the new GPU-accelerated finite-volume code \idefix\ to produce global hydrodynamics 2D simulations of the accretion discs of DN systems with a sufficiently fine spatial resolution to capture the short scale-height of cold, quiescent discs with Mach numbers ranging from 80 to 370.}
   {Running the simulations on  timescales of tens of binary orbits shows transient angular momentum transport that decays as the disc relaxes from its initial conditions. We find the angular momentum parameter $\alpha$ drops to values of $\ll 10^{-2}$, too weak to drive accretion in quiescence.}
  {}
  

   \maketitle
%

\section{Introduction}

Dwarf novæ (DNe), which are binary systems composed of a white dwarf accreting from a low-mass companion that fills its Roche lobe, display outbursts due to changes in the flow of matter through the accretion disc surrounding the white dwarf \citep{smak_eruptive_1971, osaki_accretion_1974}.  Like low-mass X-ray binaries composed of a black hole or neutron star accreting from a low-mass companion, DNe cycle between an outburst state and a quiescent state, albeit on a much shorter recurrence timescale of the order of a month.

The disc instability model (DIM; e.g. \citealt{lasota_disc_2001,hameury_review_2020} for reviews) proposes a mechanism to explain the luminosity variation of these systems, which is related to a gas-ionisation and opacity hysteresis cycle. During quiescence, the disc gas is mostly neutral and marginally optically thick ($\tau\approx 1$). During this phase, matter from the companion accumulates until a surface density threshold is reached in the disc. At this point, the strong dependence of the opacity on temperature as a result of hydrogen ionisation leads to a thermal runaway that heats the disc. Eventually, the disc ends up hot, fully ionised, and optically thick: the outburst state. In this state, the disc has a higher accretion rate onto the white dwarf than the mass fed at the outer disc from the secondary. The disc density therefore decreases until its density reaches a lower threshold where hydrogen starts to recombine. The typical recurrence time for these systems is around 40 days in total, with about a month for quiescence and a week for the outburst phase \citep{osaki_dwarf-nova_1996}.

The transport of matter is driven by the redistribution of angular momentum in the disc, historically described as a turbulent viscosity $\nu =\alpha c_{\rm s} H,$ where $c_{\rm s}$ is the local sound speed and $H$ the local scale height \cite{shakura_black_1973}. The parameter $\alpha$ depends on the physics that governs the transport of angular momentum.  Assuming a radiatively efficient thin disc, it is possible to derive the value of $\alpha$ required to reproduce the outburst cycles. \cite{mineshige_disk-instability_1983}, \cite{meyer_outbursts_1984}, \cite{smak_temperatures_1984}, and \cite{martin_physical_2019}, amongst others,  derived an $\alpha$ parameter value   of 0.1 to 0.3   during outburst and 0.01 during quiescence, based on observations. 

The magneto-rotational instability (MRI, \citealt{velikhov_stability_1959}, \citealt{chandrasekhar_hydrodynamic_1961}, \citealt{balbus_powerful_1991}) is now widely accepted as the source of turbulent angular momentum transport during the hot, outburst phase (\citealt{latter_hysteresis_2012}, \citealt{hirose_convection_2014}). However, the MRI fails to describe the angular momentum transport process during quiescence. During this colder phase, the plasma is too weakly ionised, and so MRI-driven turbulence is expected to be largely reduced \citep{gammie_origin_1998}. This conjecture has been confirmed through local shearing box models \citep{scepi_impact_2018}, which show that the molecular resistivity of such cold plasma suppresses MRI-driven angular momentum transport. Consequently, we do not yet understand the angular momentum transport mechanism that acts during the quiescent phase. 

An alternative route to angular momentum transport in MRI-stable discs is spiral shock waves. Spiral waves are known to be excited by the tidal potential of the binary, and patterns observed in the orbital light curves or Doppler tomograms of  cataclysmic variable (CV) accretion discs are consistent with their presence \citep[e.g.][]{pala_evidence_2019,ruiz-carmona_analysis_2020}. Studies using linear \citep{savonije_tidal_1983,savonije_tidally_1994,xu_linear_2018} and non-linear self-similar analysis \citep{spruit_stationary_1987,hennebelle_spiral-driven_2016} find that the angular momentum transport driven by spiral shocks is strongly correlated with temperature (typically with $\alpha\propto T^{3/2}$). This arises because the temperature controls the disc Mach number (see Sect. \ref{sec:methods}), which determines the opening angle of spiral shocks and, in turn, the angular momentum flux. Therefore, in low-temperature quiescent discs with Mach numbers of the order of several hundred we expect spiral shock to drive very weak angular momentum transport.

However, the spiral scenario was recently revived by global simulations of close binary systems \citep{ju_global_2016,pjanka_stratified_2020} that emphasise the important role of spiral density waves in the dynamics of these objects. Due to the high computational cost of global models, these simulations focused on hot discs, with Mach numbers of the order of a few tens.  Simulating high Mach numbers is challenging, as this requires high resolution to accurately capture the shallow angle between the tightly wound spiral and the Keplerian flow \citep{matsuda_mass_1990,rozyczka_numerical_1993,blondin_tidally-driven_2000,makita_two-_2000,kley_simulations_2008}. Simulations are also typically limited in the radial range and time span over which they follow the evolution of the spiral waves. It has therefore not yet been feasible to determine from direct global simulations whether or not spiral shock waves are a viable angular momentum transport mechanism in quiescence, when the Mach number can reach $900$ at the inner edge of the disc (from cold branch temperature estimates of the DIM).

In this paper, we propose to tackle this question using a new GPU-accelerated finite-volume code, \idefix, which allows us to explore these quiescent regimes. We present the first 2D global hydrodynamics model with a sufficiently fine resolution to resolve the spiral waves in a cold quiescent disc. The code is also fast enough for us to explore timescales well beyond transients due to relaxation from the initial state. In Sect. \ref{sec:methods}, we review the analytical formulation and present the code used to solve the equations. In Sect. \ref{sec:iso}, we show two-dimensional simulations of isothermal discs with realistically high Mach numbers. We discuss the consequences of our findings in Sect. \ref{sec:discussion} and present conclusions in Sect. \ref{sec:conclusion}.
  

\section{Methods\label{sec:methods}}

In this paper, we use the \idefix\footnote{Stable version and documentation can be found at \url{https://github.com/idefix-code/idefix}} \citep{lesur2023idefix} finite-volume astrophysics code to solve the hydrodynamic Euler equations in polar coordinates $(R,\phi)$. More details on the code are given in Sect. \ref{sec:algo}.

\subsection{Analytical formulation}

We solve the Euler equations for hydrodynamics:
\begin{equation}
    \frac{\partial \rho}{\partial t} + \nabla \cdot ( \rho \bm{v} ) = 0,
\end{equation}
\begin{equation}
    \frac{\partial} {\partial t} (\rho \bm{v}) + \nabla  \cdot ( \rho \bm{v}\bm{v}^T ) = - \nabla p - \rho \nabla \Psi
,\end{equation}
for a fluid of density $\rho$, velocity $\bm{v}$, with pressure $p$ and in a gravitational potential $\Psi$.
The above system is closed with the ideal gas equation of state: 
\begin{equation}
    p = c_\text{s}^2 \rho
.\end{equation}
The runs presented here are isothermal: the temperature and sound speed $c_\text{s}$ are constant over time and space. Here, since we only present 2D simulations, the volume density $\rho$ and the pressure $p$ are vertically integrated. We write these vertically integrated quantities $\Sigma$ and $P$, respectively.

We work in a rotating frame centred on the white dwarf in which the secondary star is fixed. For our binary systems, the tidal potential $\Psi$ in the rotating frame is given by
\begin{multline}
        \Psi(R,\phi) = - \frac{GM_\text{WD}}{R} - \frac{GM_\text{s}}{\sqrt{R^2 + a^2 - 2aR\cos(\phi)}} \\
        + \frac{GM_\text{s}}{a^2}R\cos(\phi),
        \label{eq:potential}
\end{multline}
where $M_\text{WD}$ and $M_\text{s}$ are the masses of the white dwarf and secondary star, respectively. $G$ is the gravitational constant, and $a$ is the binary separation\rmm{, and $\omega$ is the binary angular frequency}.  The third term of this potential accounts for the non-inertial reference frame. \rmm{At the initial time $t=0$, the companion star is set at $(x,y) = (a,0)$ and rotates counterclockwise.} The inertial \rmm{and Coriolis} forces are added by the solver. In practice, the centrifugal force is added as a radial source term $\Sigma {v_\phi^2}/{R}$ with $v_\phi$ being the total azimuthal velocity (i.e. taking both the rotation of the reference frame and a possible advection velocity). The Coriolis force is included as a modified inter-cell momentum flux to guarantee angular momentum conservation at machine precision \citep{mignone_conservative_2012}.

\subsection{Grid and units}

For all simulations in this paper, we use a logarithmic grid in the radial direction, such that $\Delta R \propto R$ and a uniform grid in the azimuthal direction. The former allows a finer resolution of the inner edge of the disc, where the spiral shock structures are smaller \citep{spruit_stationary_1987}. We set the inner boundary of the disc at $r_0 = 0.01 \, a$, which corresponds to the expected radius of the white dwarf. The outer radius of the integration domain is chosen to be the radial position of the Lagrange point $L_1$ of the binary system. With these definitions, the\rmm{inner and} outer radius changes with the mass ratio $q=M_\text{s}/M_\text{WD}$. However, we choose the resolution such that going to a further radius only appends points to the grid, with no modification of the inner region (this concerns the runs $\mathrm{Ma} = 250$ from $q=0.3$ to $q=0.1$ and $q=0.7$).

In this paper, we choose the time unit as the binary period $T_0 = \frac{2\pi}{\omega} = 2\pi \sqrt{\frac{a^3}{G (M_\text{s}+M_\text{WD})}}$. The length unit is the binary separation $a$. Following \citet{ju_global_2016}, these units are scaled on the dwarf nova system SS Cygni, where $T_0$ is approximately 6.6 hours and $a$ is 1.37$\times10^{11}$ cm \citep{bitner_masses_2007}.

We define the Mach number at the inner edge as 
\begin{equation}
    \mathrm{Ma} = \frac{\Omega_K(r_0) r_0}{c_\text{s}}\approx 364 \left(\frac{M_{\rm WD}}{\text 1\,M_{\odot}}\right)^{1/2}\left(\frac{10^{9}\rm\,cm}{r_{0}}\right)^{1/2}\left(\frac{10^{4}\rm\,K}{T}\right)^{1/2}
    \label{eq:Mach}
,\end{equation}
where $\Omega_K(R) = \sqrt{{GM_\text{WD}}/{R^3}}$ is the Keplerian angular frequency around the white dwarf. We note that this definition depends on the chosen inner boundary of the simulation; for example, in an isothermal setup $\mathrm{Ma} \propto \sqrt{r_0}$. This must be taken into account when comparing with previous work. In particular, \citet{ju_global_2016} set their inner boundary further out at  $r_0=0.02 \, a$.

\subsection{Algorithm\label{sec:algo}}

The simulations presented in this paper all use the \idefix\ code, which is a finite-volume conservative Godunov grid-based code with a structure similar to {\it Pluto} \citep{mignone_pluto_2007}. The code is written in C++17 and uses the Kokkos portability library \citep{carter_edwards_kokkos_2014} for many-core shared memory parallelisation and MPI for distributed memory parallelisation, allowing for high performance on most available architectures. 

We use the HLLC Riemann problem solver \citep{harten_upstream_1983, toro_restoration_1994}. We also use the Fargo algorithm \citep{masset_fargo_2000} with a Keplerian advection velocity to speed up the integration, as implemented in \citet{mignone_conservative_2012}. We verified that we are able to reproduce the work of \citet{ju_global_2016} with this code (see details in Appendix \ref{app:ju}).

\begin{table}
  \caption[]{Runs presented in this paper.}
  \label{tab:runs}
  \centering
 $$
     \begin{array}{ccc}
        \hline
        \noalign{\smallskip}
        \mathrm{Ma}(r_0)      & \text{Resolution } (N_R \times N_\phi) & q \\
        \noalign{\smallskip}
        \hline
        \noalign{\smallskip}
        
                80 &  1081 \times 1024 & 0.3\\
                140 & 1081 \times 1024 & 0.3\\
                250 & 1081 \times 1024 & 0.3\\
                370 & 1081 \times 1024 & 0.3\\
                \hline
                250 & 1119 \times 1024 & 0.1\\
                250 & 1042 \times 1024 & 0.7\\
                250 \textrm{WKZ} & 1081 \times 1024 & 0.3\\
                550 \textrm{HR} & 4096 \times 4096 & 0.3\\

                \noalign{\smallskip}
        \hline
     \end{array}
 $$
 \end{table}

\subsection{Boundary conditions and density floor}
In the azimuthal direction, we use periodic boundary conditions while we use `Keplerian outflow' boundary conditions in the radial direction. These last boundary conditions are identical to standard `outflow' boundaries, with the exception of the azimuthal velocity, which is set to its Keplerian value in the ghost zones. We additionally use a density floor to limit the density contrast in the simulation. If the density of one cell drops below the threshold $\Sigma_\mathrm{floor} = 10^{-6}$, then this value is used as density instead. The density floor is implemented in a total momentum conserving fashion, so as to prevent the injection of angular momentum by the procedure.  

\subsection{Initial conditions}

We take the initial profile of the disc to have a uniform surface density $\Sigma = 1$. The initial velocity profile is Keplerian with $v_\phi =R \Omega_\text{K} =\sqrt{GM_\text{WD}/R}$ and no initial radial velocity. This state is not an equilibrium state, because the initial velocity only takes into account the gravitational influence of the central white dwarf, not the companion.  

We pretruncate the disc at the outer edge to shorten the time needed for the outer disc to settle down. The initial truncation radius is chosen to be slightly larger than the expected $r_\text{max}$ of \cite{paczynski_model_1977}.  The truncation is performed using the following mask applied to the initial density profile:
\begin{equation}
   \mu(R) =  \frac{1}{2} \left ( 1 - \tanh \left ( \frac{R - r_T}{\delta_T} \right ) \right),
\end{equation}
where $r_T$ is the chosen truncation radius and $\delta_T=0.01$ is the chosen truncation width. We note that this truncation is not perfect as it is axisymmetric, unlike the predicted last stable disc orbit \citep{paczynski_model_1977}. At larger radii, we enforce the previously described density floor.  We typically wait for the system to relax over a few binary orbital periods before measuring any quantity.

\subsection{Averaging methods}

In this work, the dynamical timescale is $\simeq $ 1/10,000$^\text{th}$ of the binary period, while we are following the evolution of the system for dozens of binary orbits. To capture the evolution in a computationally efficient way, we used an `on-the-fly' time and azimuthal averaging scheme: we sample the simulation every 1/10,000$^\text{th}$ of a binary orbit and average over 100 of these samples, producing an output each 1/100$^\text{th}$ of a binary orbit. This is sufficient to capture the rapid inner-edge dynamics.

\section{Results\label{sec:iso}}
In this section, we present globally isothermal models with realistically low temperatures (high Mach numbers). 

\subsection{Resolution}
We chose the resolution such that the expected spiral wave is well resolved everywhere in the disc. Linear theory predicts that the wavelength at the inner edge of the disc is \citep{savonije_tidally_1994} 
\begin{align}
    \label{eq:lambda}
    \frac{\lambda}{R} &\simeq \frac{2\pi}{\sqrt{m^2-1}}\sqrt{\frac{R}{r_0}}\mathrm{Ma}^{-1}(r_0),
\end{align}
where $m$ is the azimuthal wave number. For the $m=2$ mode that is expected to be dominant \cite{savonije_tidally_1994}, a resolution of $N_R = 1000$ gives a minimum of 2.5 radial points per wavelength for our maximum Mach number $\mathrm{Ma}=370$ (Fig. \ref{fig:resolution}). We consider this sufficient to resolve the entire spiral in the disc.  We use about as many points in the azimuthal direction as in the radial direction to keep a cell aspect ratio of $\frac{Rd\phi}{dR} \simeq 1.6$, that is, of order unity. Therefore, the resolution of our isothermal runs is 1081 radial points and 1024 azimuthal points (Table \ref{tab:runs}). The run with $\mathrm{Ma}=250$ and $q=0.1$ has a larger radial extension and consequently larger $N_R$ to keep  the grid spacing unchanged in the inner disc. The opposite is true for the run with $q=0.7$. 

   \begin{figure*}
            \begin{subfigure}{\columnwidth}
                \includegraphics[width=\textwidth]{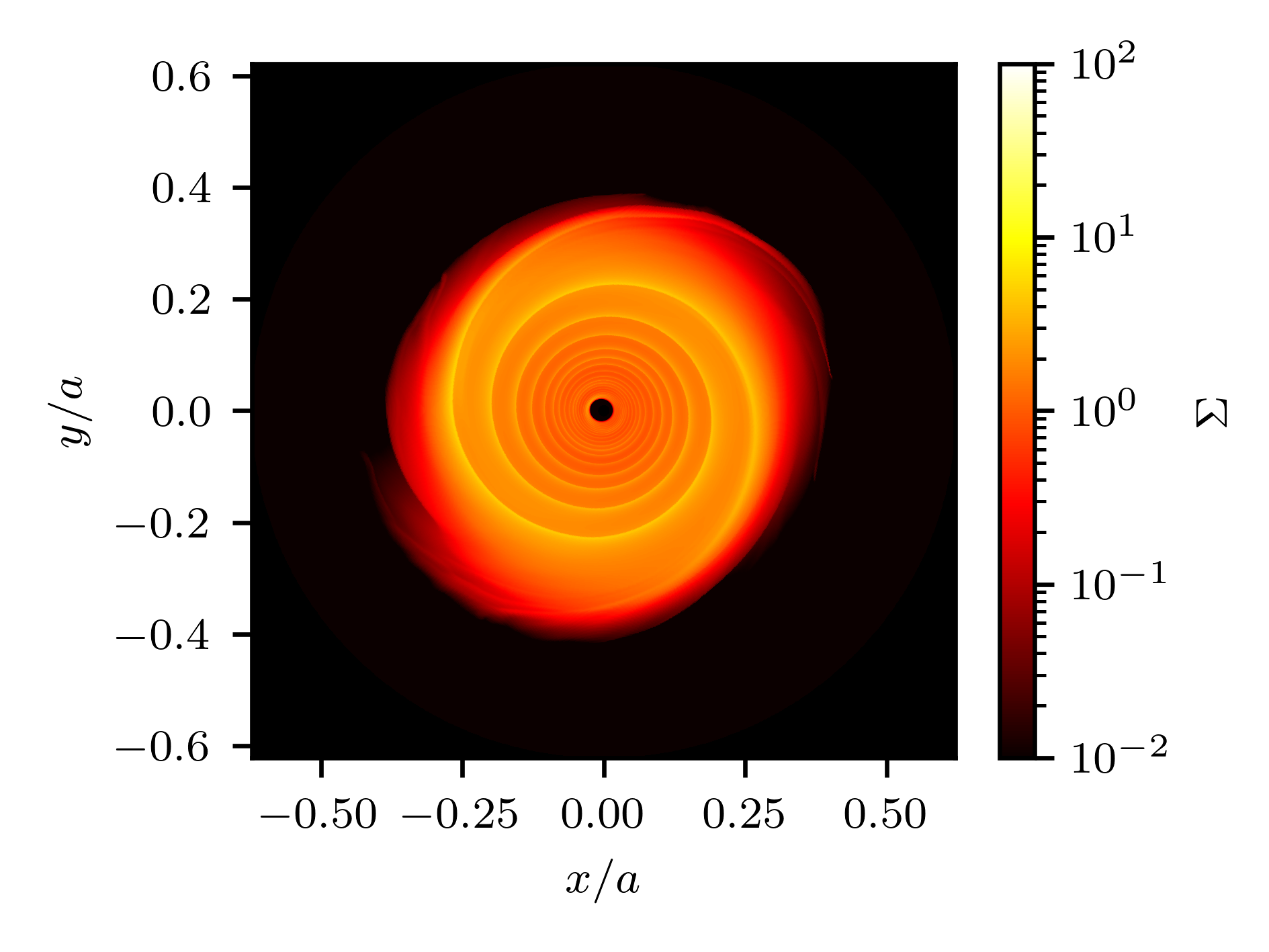}
         \caption{$\mathrm{Ma} = 80$}
              \end{subfigure}
              \begin{subfigure}{\columnwidth}
                \includegraphics[width=\textwidth]{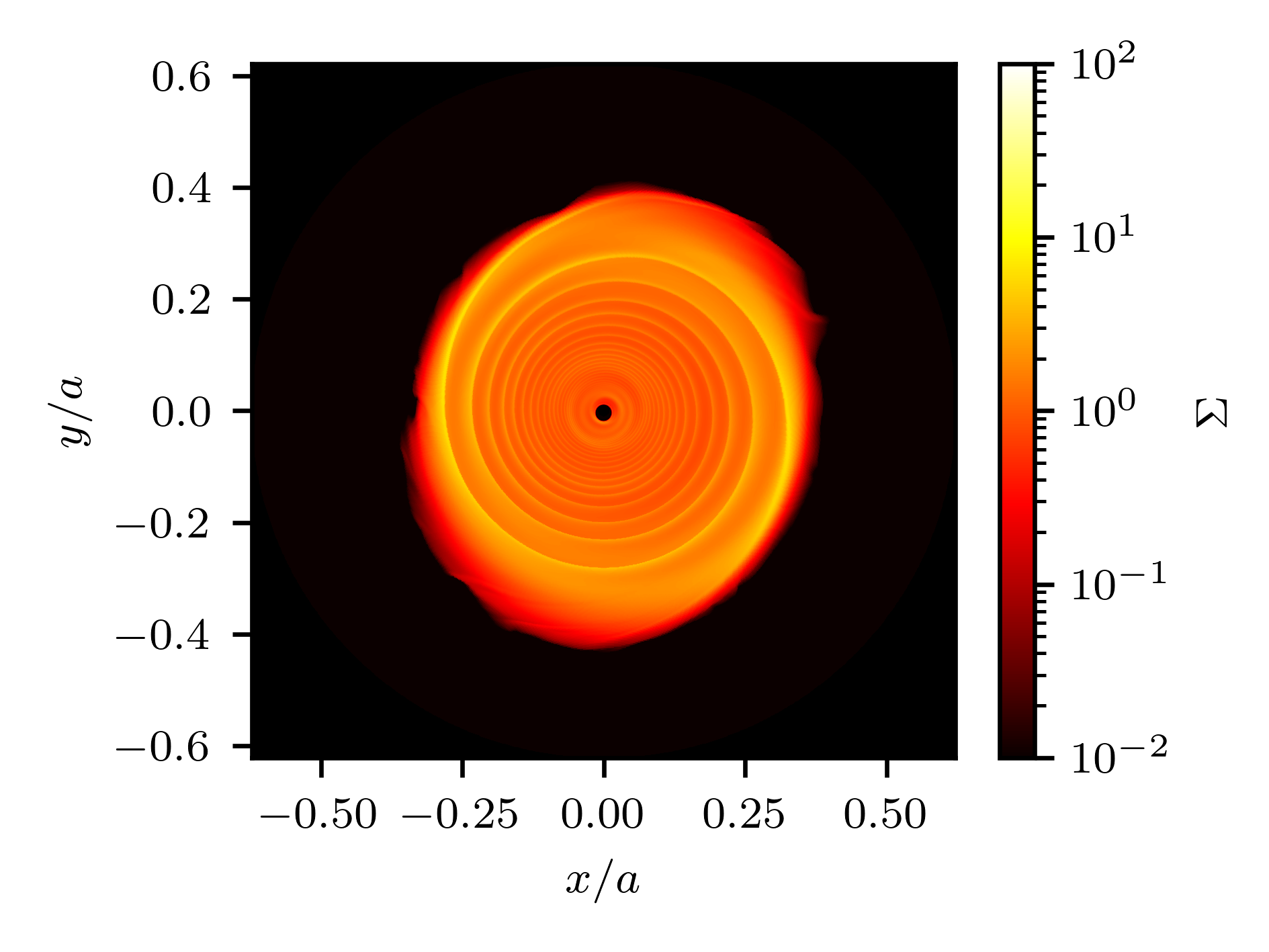}
         \caption{$\mathrm{Ma} = 140$}
              \end{subfigure}
              \\
              \begin{subfigure}{\columnwidth}
                \includegraphics[width=\textwidth]{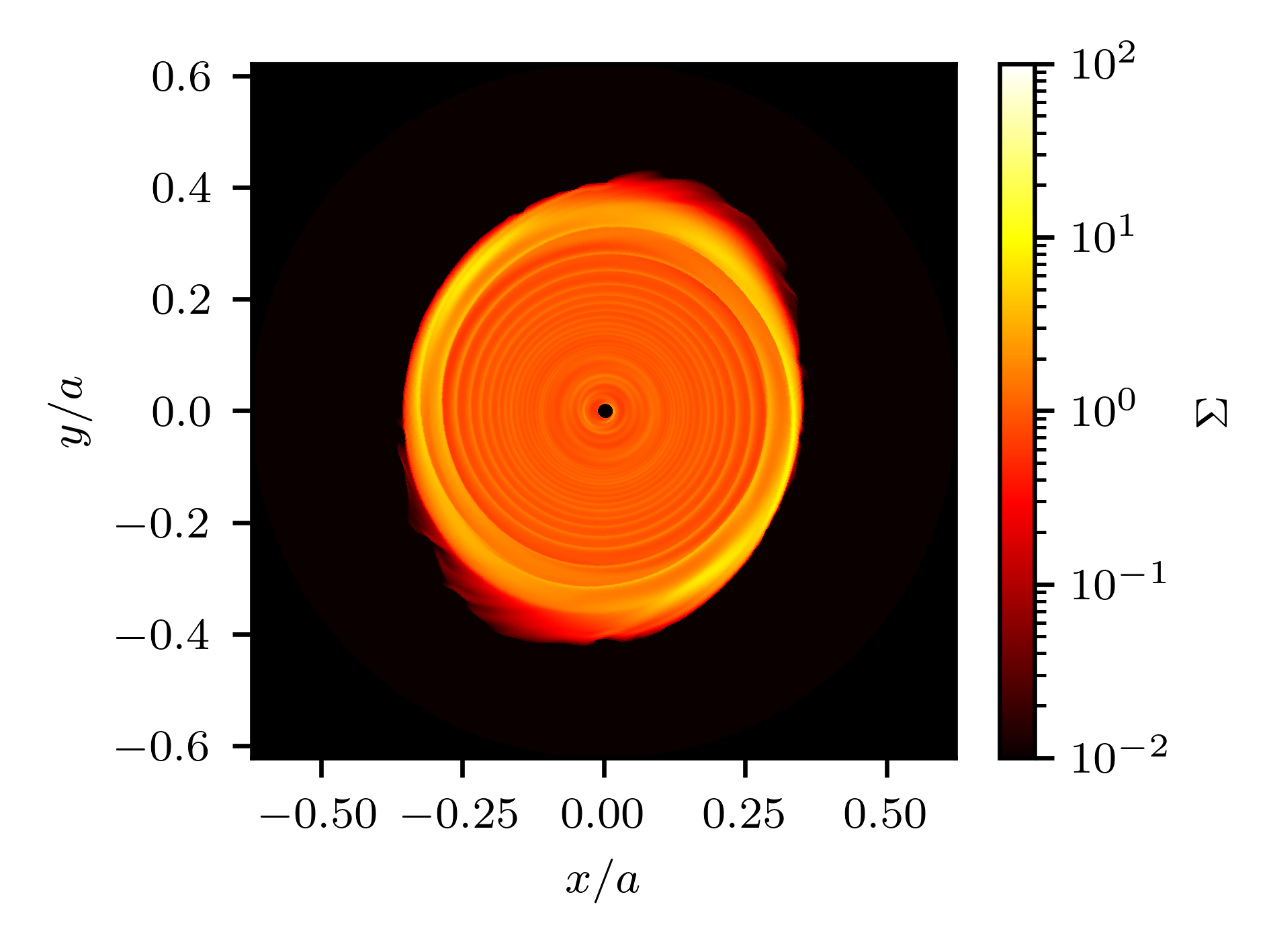}
         \caption{$\mathrm{Ma} = 250$}
              \end{subfigure}
              \begin{subfigure}{\columnwidth}
                \includegraphics[width=\textwidth]{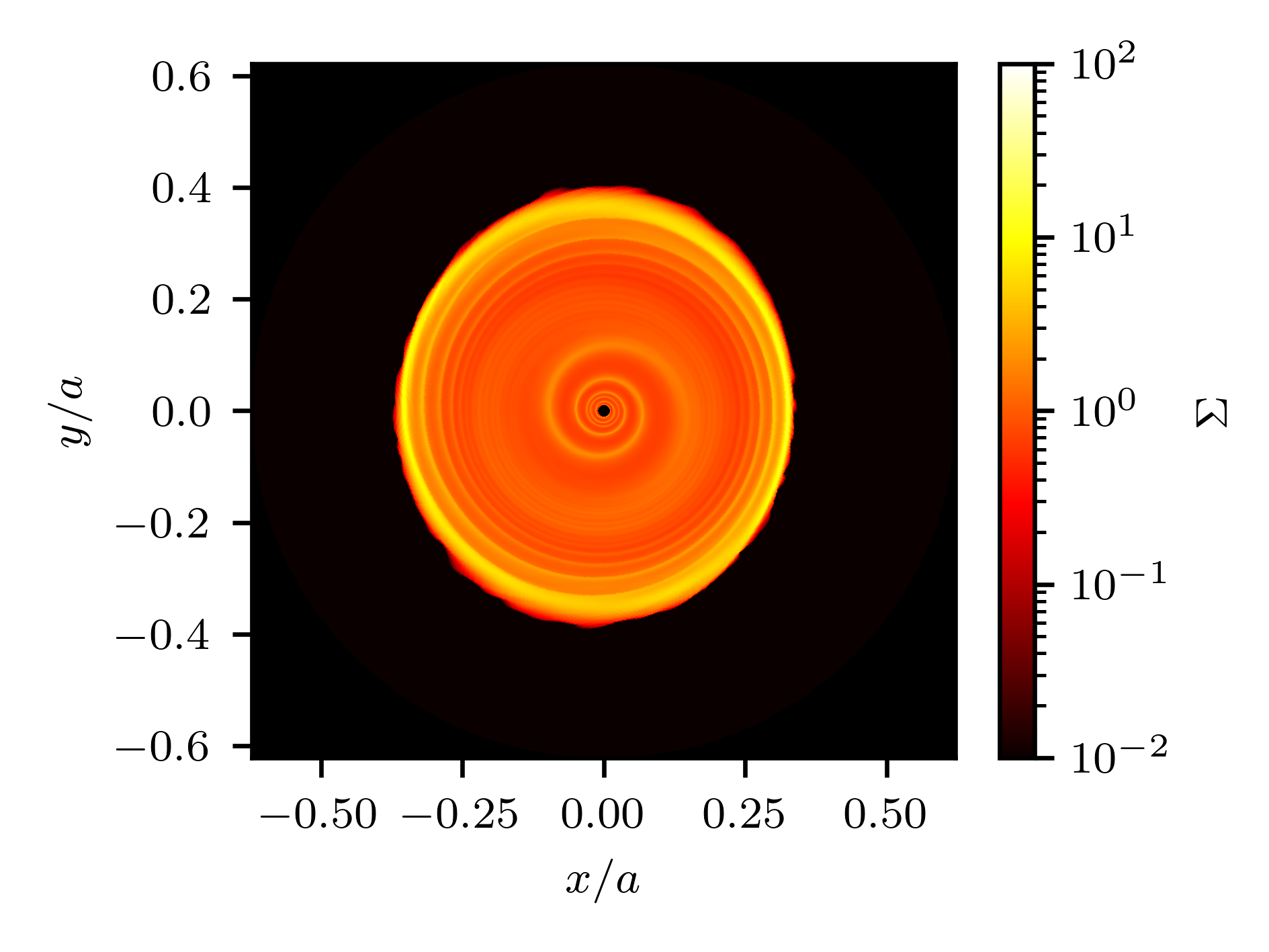}
         \caption{$\mathrm{Ma} = 370$}
              \end{subfigure}
        \caption{Density maps after ten binary orbits for isothermal runs with different Mach numbers.}
         \label{fig:MapDensity}
   \end{figure*}
   
    \begin{figure*}
            \begin{subfigure}{\columnwidth}
                \includegraphics[width=\textwidth]{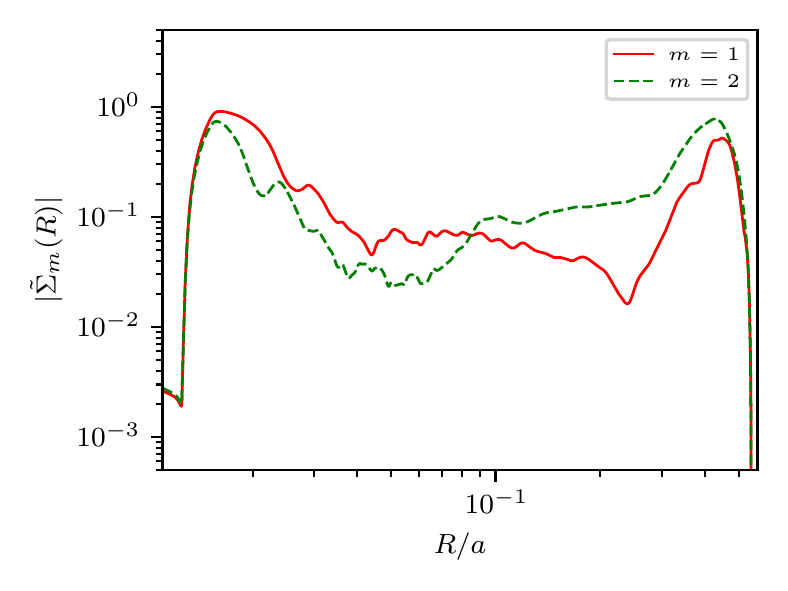}
         \caption{$\mathrm{Ma} = 80$}
              \end{subfigure}
              \begin{subfigure}{\columnwidth}
                \includegraphics[width=\textwidth]{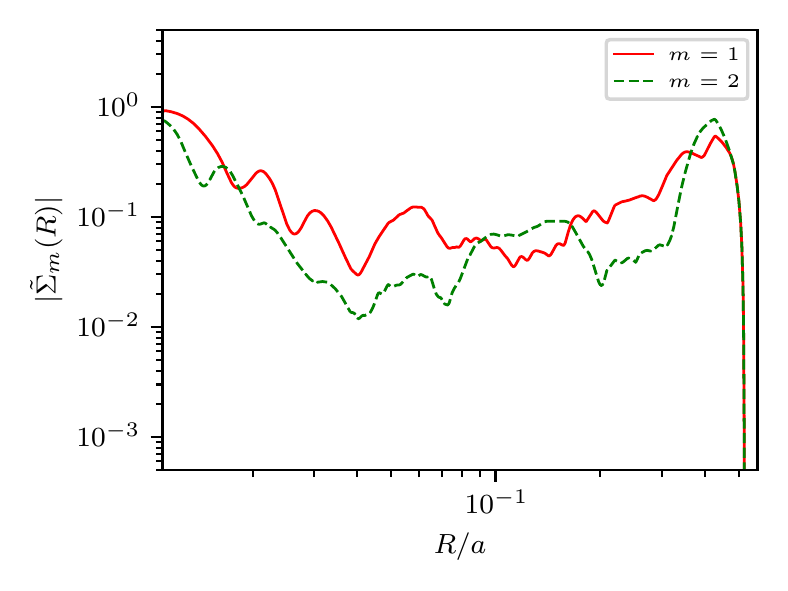}
         \caption{$\mathrm{Ma} = 140$}
              \end{subfigure}
              \\
              \begin{subfigure}{\columnwidth}
                \includegraphics[width=\textwidth]{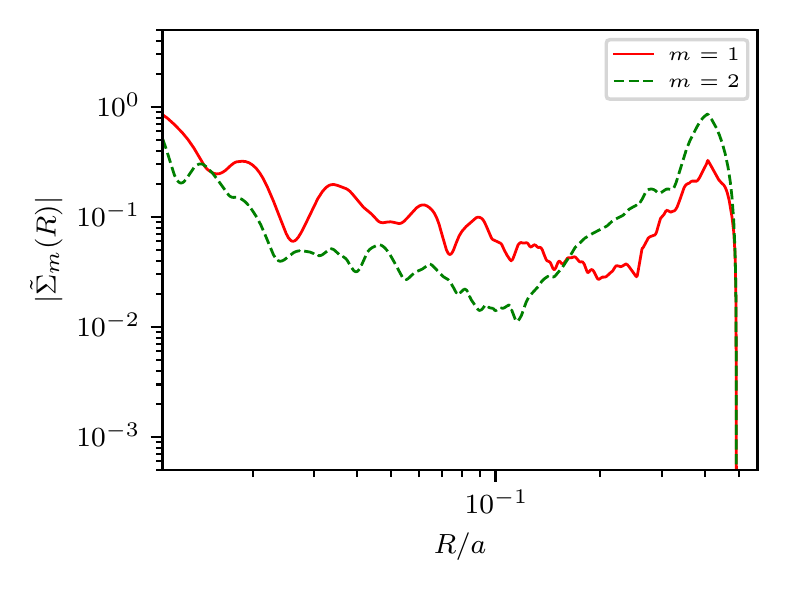}
         \caption{$\mathrm{Ma} = 250$}
              \end{subfigure}
              \begin{subfigure}{\columnwidth}
                \includegraphics[width=\textwidth]{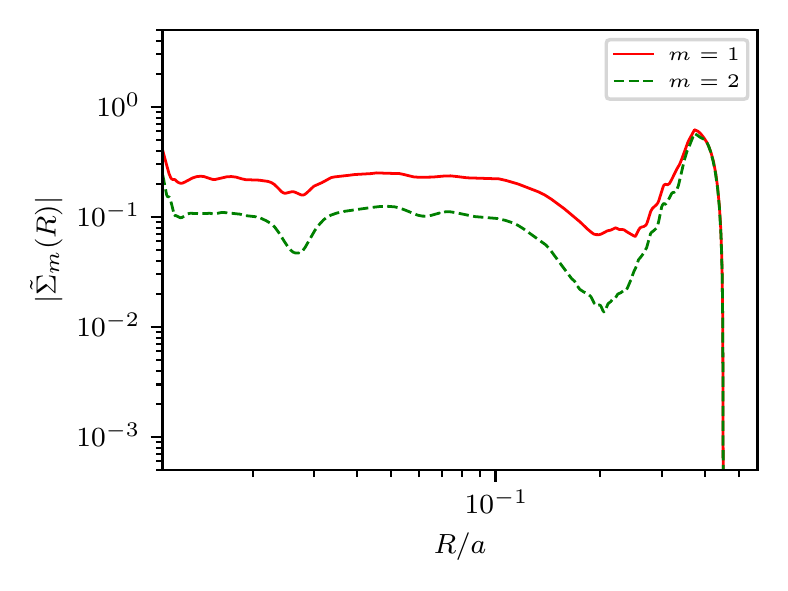}
         \caption{$\mathrm{Ma} = 370$}
              \end{subfigure}
        \caption{Azimuthal spectrum of the relative  density fluctuation distribution for modes $m=1$ (red solid line) and $m=2$ (green dashed line) after ten binary orbits. }
        \label{fig:DensityTF}
   \end{figure*}

\subsection{Spiral waves}
\label{sec:spiral}
As expected from previous works (Sect. 1), we see that spiral waves develop in the accretion disc due to the tidal potential (Fig. \ref{fig:MapDensity}). 
To get a better grasp of the excited modes, we show the Fourier decomposition of the density in Fig. \ref{fig:DensityTF}. The plotted values correspond to the amplitude of the Fourier expansion of the relative density fluctuation $\vert \tilde{\Sigma}_m(R) \vert $ smoothed over 50 radial points (moving average), defined from
\begin{equation}
    \tilde{\Sigma} (R,\phi) = \frac{\Sigma(R,\phi) - \langle\Sigma\rangle_\phi(R)}{\langle\Sigma\rangle_\phi(R)} = \sum_{m=0}^\infty  \tilde{\Sigma}_m(R) e^{2i\pi m\phi},
    \label{eq:TF}
\end{equation}
where $\langle \Sigma \rangle_\phi(R) = \frac{1}{2\pi} \int_0^{2\pi} \Sigma(R,\phi) d\phi$ is the azimuthal average of the surface density. Figure \ref{fig:DensityTF} shows strong $m=1$ and $m=2$ spiral modes are present in all runs. The $m=2$ mode predicted by linear theory dominates only at low Mach numbers and in the outer disc. The simulations show that the $m=2$ mode is initially excited at the outer boundary, propagates through the disc, and is reflected when it reaches the inner edge. The strong $m=1$ mode becomes visible in density maps after this reflection together with the formation of an eccentric inner cavity. 

To investigate the impact of the chosen inner boundary conditions on the appearance of the $m=1$ mode, we performed an additional run at $\mathrm{Ma} = 250$ and $q=0.3$ implementing a wave-killing zone close to the inner edge extending from $r_0$ to  $r_\text{WKZ}$. In this zone, we relax all physical quantities to the initial Keplerian state of the disc, with constant surface density and no radial velocity. The associated relaxation timescale is set to one-tenth of the orbital time $\Omega_K^{-1}(r_0)$ at the inner boundary. For a wave-killing zone to be effective, one needs to choose its spatial extension to be a few wavelengths of the mode one wishes to neutralise (see figure \ref{fig:resolution}). In our case, we first take $r_\text{WKZ} = 1.2 r_0$, so that the $m=2$ spiral mode is cancelled at the inner edge. However, this had no significant impact either on the dynamic\rmm{,} or on the eccentric cavity at the inner edge. Using a more extended wave-killing zone with $r_\text{WKZ} = 2 r_0$ yields the same results. We conclude that there is more at play than a mere reflection exciting the $m=1$ mode, most likely because we do not fall within the linear theory regime (see Sect.\,\ref{sec:discussion}).

\subsection{Measured $\alpha$}
\label{sec:measuredalpha}

We measure the effective angular momentum transport parameter $\alpha$ \citep{shakura_black_1973} in our simulations as
\begin{equation}
    \alpha(R) = \frac{\langle F_R(v_\phi) \rangle_{t,\phi}}{\langle \rho c_\text{s}^2\rangle_{t,\phi}},
\label{eq:alpha}
\end{equation}
where $F_R(v_\phi)$ is the inter-cell angular momentum flux in the radial direction $R,$ which can be written $F_R(v_\phi) = \rho v'_r v'_\phi$, where $v'_\phi = v_\phi - v_\text{K}$ is the departure from the Keplerian velocity. The averages are computed over the whole azimuthal domain $\phi \in \left [ 0 , 2\pi \right ]$, and over a few binary orbits. The exact time span over which the average is carried out will always be specified in the following. We use the inter-cell angular momentum fluxes computed by the Riemann solver in the integration loop rather than the reconstructed cell-centred values. This allows us to better capture the small-scale dynamics. We compare our method to measure $\alpha$ with that of  \cite{ju_global_2016} in Appendix \ref{app:alpha}.

\begin{figure*}[h]
\centering
\includegraphics[width=\hsize]{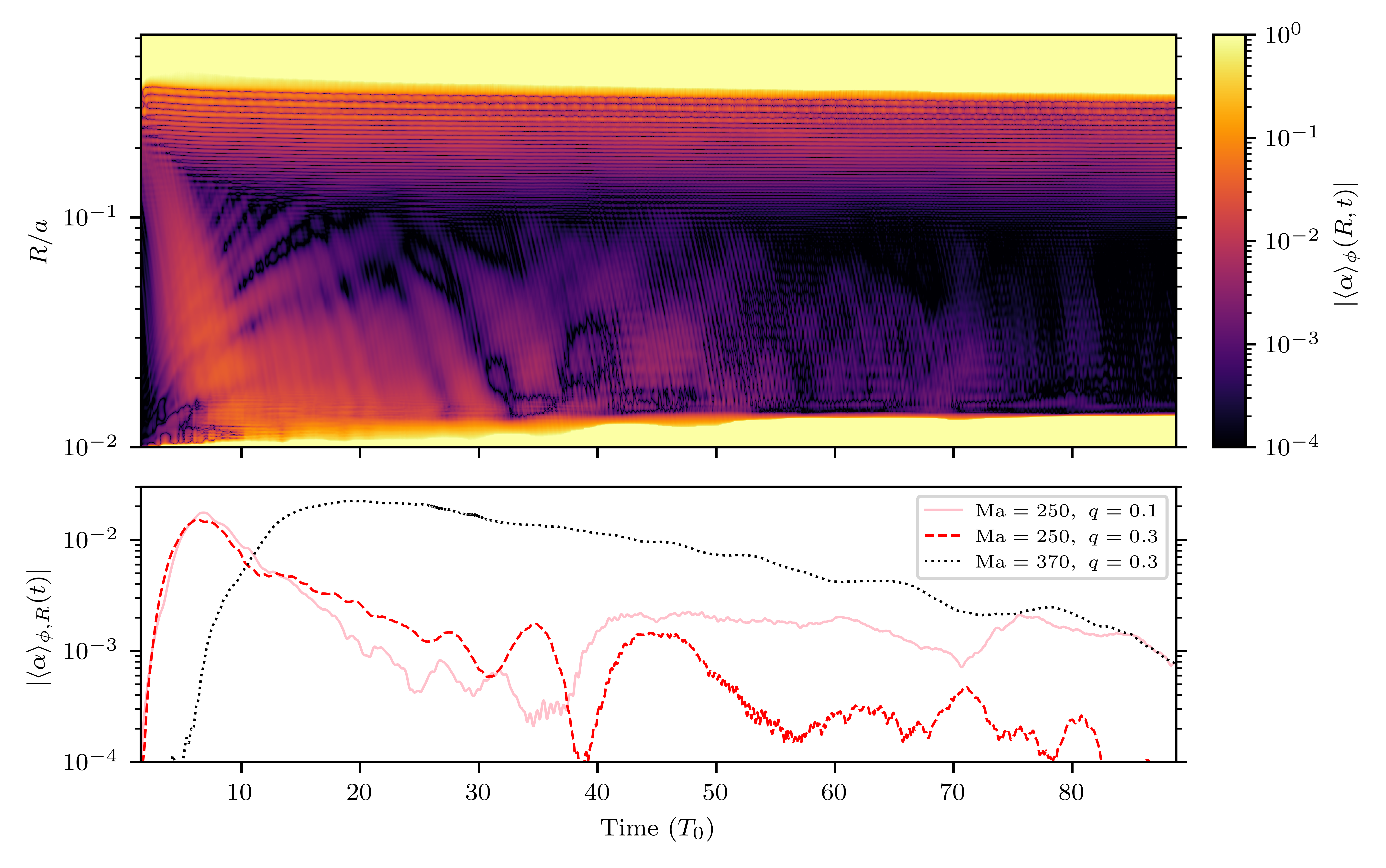}
  \caption{Measured angular momentum transport parameter. Top panel: Measured angular momentum transport parameter $\alpha$ azimuthally averaged and smoothed over three binary orbits for the run $\mathrm{Ma} = 250$ and $q=0.3$. Here we plot the absolute value of the transport parameter, and note that its sign changes in the spiral waves (oscillations at $R\gtrsim 0.1$) and is mostly positive everywhere else. Bottom panel: Azimuthally and radially (over $0.02 \leq R \leq 0.1$) averaged $\alpha$, smoothed over three binary orbits for three runs.}
     \label{fig:KP2D}
\end{figure*}

Figure \ref{fig:KP2D} shows the time evolution of $\alpha$ in the disc. In all simulations, we first see a transient regime at early times ($t\lesssim 20$) with a high $\alpha$ that slowly decays. At later times ($t\gtrsim 40$), the transport parameter in the disc drops to $\alpha \ll 10^{-2}$ in all simulations, regardless of the mass ratio $q$ or the Mach number. The relaxation timescale appears to be dependent on the Mach number. For example, the simulation with $\mathrm{Ma} = 370$ shows a slower relaxation than the $\mathrm{Ma} = 250$ run (bottom panel of Fig. \ref{fig:KP2D}). The state reached after 80 orbits may still not be a steady state and $\alpha$ is still decreasing. We did not go beyond this stage, as $\alpha$ is already much lower than required by the DIM. 

We caution that regions of very high $\alpha$ at the inner and outer edges are unphysical regions where the density floor is triggered because of the  growth of an eccentric inner cavity (Sect. \ref{sec:spiral}) or the tidal truncation of the outer disc. We also note that matter accumulates at the outer radius of the disc at lower temperatures. This is a result of relaxation of the disc truncation from its initial state, which assumes a Keplerian velocity profile and an analytical guess for the outer radius. This residual matter is not efficiently redistributed by the spiral waves given the weaker angular momentum transport at lower temperatures.

\begin{figure}
\centering
\includegraphics[width=\hsize]{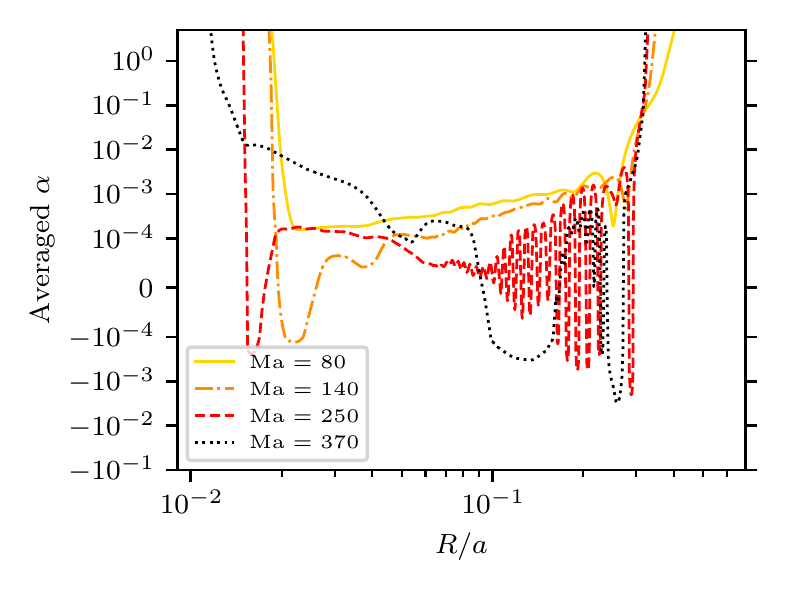}
  \caption{Measured angular momentum transport parameter $\alpha$ at late times (between $t=80$ and $t=89.9$) for different Mach numbers with $q=0.3$, averaged over time and azimuth.}
     \label{fig:KPlate}
\end{figure}

Figure \ref{fig:KPlate} shows the radial distribution of $\alpha$ at late times for simulations with different Mach numbers, keeping $q=0.3$. The plotted values are both azimuthally- and time-averaged. The strong oscillations around $R=0.2-0.3$ reflect the effect of the tidal torque on the spiral waves. Here, unlike in \cite{ju_global_2016}, the torque term of the angular momentum conservation equation does not exactly balance the Reynolds stress term. As a consequence, the measured accretion rate and the measured $\alpha$ are not exactly zero, but rather oscillate, which creates regions of slightly negative accretion rate. The efficiency of transport decreases with disc temperature, and in all cases we get $\alpha \ll 10^{-2}$. We do not find evidence for a radial steepening of $\alpha$ with higher Mach numbers \citep[e.g.][]{blondin_tidally-driven_2000}.

\begin{figure}
\centering
\includegraphics[width=\hsize]{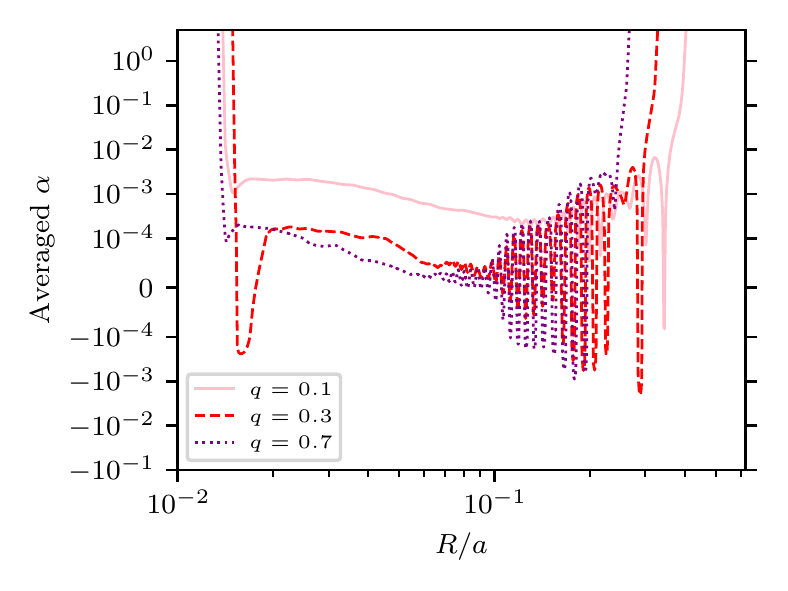}
  \caption{Measured angular momentum transport parameter $\alpha$ at late times (between $t=80$ and $t=89.9$) for different mass ratios with $\mathrm{Ma}=250$, averaged over time and azimuth.}
     \label{fig:KPq}
\end{figure}

Figure \ref{fig:KPq} shows how $\alpha$ changes with varying mass ratio $q$, keeping $\mathrm{Ma}=250$. Again, we obtain $\alpha \ll 10^{-2}$ at late times. However, unlike the runs at $q=0.3$ and $q=0.7$, the transport appears to stabilise at a higher value of $\alpha\approx 2\times 10^{-3}$ for $q=0.1$. This is consistent with a possible coupling with the 3:1 resonance, which is inside the disc for $q<0.3$ \citep{lubow_model_1991}.

\subsection{Convergence study}

In order to make sure that the resolution of our runs is sufficiently fine, we performed a higher-resolution run of our coolest setup. For this simulation, we increase the resolution by a factor of four, with a grid of size $4096\times 4096$. This resolution is much finer than what is required to resolve the $m=2$ spiral wave from linear theory. However,  the theoretical estimate may not hold given that we are far from the linear regime (see Sect.\,\ref{sec:discussion}). We observe the same spiral wave structure in the high-resolution run as in the low-resolution run, albeit with a slower decrease in $\alpha$. This is likely a consequence of the decreased numerical diffusivity at higher resolution. The relaxation timescales that we observe in our runs (Fig.~\ref{fig:KP2D}) are therefore partly controlled by the numerical diffusivity. We reach $\alpha\ll 10^{-2}$ after 100 binary orbits, a value consistent with the lower-resolution runs, confirming that these low values of $\alpha$ are not an artefact of resolution.

\section{Discussion\label{sec:discussion}}
\subsection{One-armed spirals}

Our simulations show that the $m=1$ modes are excited concomitantly with $m=2$ spiral waves\rmm{on the inner boundary}.  Previous works indicated that $m=1$ modes could be excited in a close binary system provided (1) that the mass ratio is low enough ($q<0.3$) for the 3:1 resonance radius to lie inside the disc and (2) that the disc is viscous, so as to couple the $m=2$ and $m=3$ modes to excite the $m=1$ modes  \citep{heemskerk_hydrodynamic_1994,stehle_hydrodynamic_1999,kornet_hydrodynamical_2000,kley_simulations_2008}. Here, we satisfy neither of these criteria: we do not include viscosity and our results hold for both $q=0.3$ and $q=0.7$. We do not observe a strong global eccentricity growth of the disc as  \citet{kley_simulations_2008} do: the eccentricity mode is limited to the inner disc region, keeping the overall shape of the outer disc circular (see Appendix \ref{app:pitch}). The emergence of this inner eccentric cavity may be of relevance to observations that suggest the presence of a truncated disc (e.g. \citealt{balman_x-ray_2012}).

Our results also hold when including a wave-killing zone at the inner boundary to dampen the $m=2$ reflection. This suggests a less efficient, non-linear coupling is at play in our simulations that may have been missed in lower-resolution simulations with lower Mach numbers. Following the analytical development from \cite{savonije_tidally_1994}, we can show that linear theory is expected to fail for the realistic temperature regimes that we explore here. From the general solution derived in this latter paper,  the amplitude of the density perturbation of the solution $\Sigma'$ compared to the unperturbed density profile $\Sigma_0$ can be quantified as
\begin{equation}
    \sigma (r) = \frac{\Sigma'}{\Sigma_0} = \frac{3}{4} \frac{1}{1+q}\left ( \frac{\omega}{\Omega_K} \right )^2 \text{Ma}^2,
\end{equation}
where $\omega$ is the binary angular frequency. For a quiescent disc temperature of 3,000 K and with typical masses for the white dwarf and its companion star, we obtain $\sigma(r_0) \simeq 1$ at the inner edge. As $\sigma(r) \propto r^2$, non-linear effects increase at larger radii. The solution derived by \cite{savonije_tidally_1994} is based on a perturbative WKB solution where $\sigma \ll 1$ and with the assumption that $r\ll a$. The emergence of a one-armed spiral wave could not have been predicted by linear theory, as the regime probed here strongly deviates from this approximation.

\subsection{Angular momentum transport}

The initial high value of the $\alpha$ parameter mentioned in Sect. \ref{sec:measuredalpha} is likely a consequence of our initial condition, which assumes circular Keplerian orbits for the gas. This is clearly not an equilibrium when a secondary star is present. Relaxation from these conditions takes more than 20 orbits. This is longer than for example the time over which \citet{ju_global_2016}  ran their simulations, meaning that their values of $\alpha$ may still be impacted by their initial conditions (which are similar to ours). However, the relaxation time is only a few binary orbits at the lower Mach numbers of their simulations.
 
 On long timescales, our simulations show that $\alpha$ is always much smaller than the typical $\alpha \approx 0.01$ in quiescence required by the comparison of DIM models to observed light curves. The values we reach are of the order of $\alpha\approx 10^{-4}$ and are still decaying when we end the simulation. The one exception is for $q=0.1$ where transport stabilises to a larger $\alpha\approx 10^{-3}$, probably because the 3:1 resonance allows waves inside the disc to be excited. This can only apply to systems with $q<0.3$, such as the WZ Sge subclass \citep{kato_wz_2015}, while DNe show mass ratios in the range of $0.1\lesssim q \lesssim 0.8$  \citep{otulakowska-hypka_statistical_2016}.

 In our simulations with the  highest Mach number, the relaxation time is comparable to the total integration time of 100 binary orbits, which is on the order of the recurrence time between two normal outbursts in DNe systems \citep{otulakowska-hypka_statistical_2016}.  Spiral waves therefore might still play a role in transporting angular momentum in quiescence if the relaxation to low $\alpha$ is slow enough. There are two caveats that make this scenario unlikely. First, our convergence run indicates that the relaxation time is likely to depend on the numerical resolution, and so deducing a reliable decay timescale from the simulation is not straightforward. Second, this would depend critically on how the disc relaxes from MRI-driven transport in outburst. Our initial conditions assume an isothermal disc with a flat surface density profile. A constant-temperature disc is a reasonable approximation for a DIM quiescent disc. However, the surface density profile is expected to increase with radius ($\Sigma \propto R$) when the disc enters quiescence due to the dependence on radius of the critical densities associated with the temperature hysteresis \citep{lasota_disc_2001}. This may introduce differences with our flat $\Sigma$ profile and, more generally, induce a complex interaction in the time evolution of the $\alpha$  and $\Sigma$ density profiles.

\subsection{RWI/KHI spiral excitation}

We also verified that the spiral-driven accretion that we observe is indeed a consequence of the presence of a secondary star. \cite{lesur_spiral-driven_2015} showed that an infalling envelope around the disc could trigger a mixture of a Rossby Wave instability \citep{lovelace_rosby_1999} and centrifugal instability at the disc outer edge, creating spiral perturbations that propagate inwards. The outer edge of our disc fulfils the instability criteria of these two instabilities, in part because of the density floor that induced an inflow from the `void' surrounding the disc, but also because of the sharp truncation caused by the secondary.

To check the impact of these edge instabilities, we simulated the same discs without the companion, keeping the disc truncated density profile. We find that in this case, these instabilities only drive transport $\alpha \ll 10^{-4}$ in the disc bulk and are therefore of negligible importance in the present work (see Fig. \ref{fig:NoSalpha}).

However, it is possible that the tidal potential of the secondary star selects and excites a specific mode of the edge instabilities. This could be a possible mechanism to explain the unexpected apparition of $m=1$ spiral modes.

\subsection{Expected instabilities in 3D}

Several instabilities expected in 3D are absent from our 2D approach, among which the vertical shear instability  \citep[VSI,][]{nelson_linear_2013} and the convective over-stability \citep[COV,][]{Klahr2014}. The existence of these two instabilities is mostly controlled by the dimensionless cooling timescale $\Omega_K \tau_\mathrm{cool}$ where $\tau_\mathrm{cool}$ measures the typical thermal relaxation timescale resulting from the heating/cooling equilibrium. In the case of CVs, we expect $\Omega_K \tau_\mathrm{cool}\ll 1$, a regime in which we expect the VSI to be present but not baroclinic instabilities \citep{Lesur2022}.

The question is then how much angular momentum transport is to be expected from the VSI. Because the VSI is driven by the vertical shear of the flow, it depends strongly on the disc thickness. Using high-resolution 3D hydrodynamical simulations, \cite{Manger2020} proposes a scaling for the VSI $\alpha\lesssim (H/R)^{2.6}$, which implies $\alpha<10^{-5}$ in a quiescent DNe disc subject to VSI turbulence. This means that VSI turbulence, if present, is probably negligible as an angular momentum transport process in these systems.

\subsection{Accretion stream\rmm{er}\ impact and thermodynamics}

The contribution of the matter flux coming from the companion star (called accretion stream\rmm{er}) is expected to be the strong heating at the `hot spot' where it meets the disc. As a consequence, we did not include it in this isothermal study.
Modelling the thermodynamics of the disc would allow us to know the heating efficiency of the spiral shocks presented in this work. Previous works such as that of \cite{ju_global_2016} showed that without any cooling, and using $\gamma=5/3,$ the disc heats up quickly. This suggests that some parts of the spiral-heated disc might still be unstable to MRI even during quiescence.

\section{Conclusion\label{sec:conclusion}}
We present the first 2D hydrodynamics simulations of a DN disc in realistic quiescence temperature regimes. Our main results can be summarised as follows.
\begin{enumerate}
    \item The dominant spiral mode is a one-armed $m=1$ mode once the disc has settled. Linear theory, which predicts an $m=2$ mode, fails at low temperatures (high Mach number), as expected.
    \item Spiral-wave-driven accretion in realistic temperature regimes is much lower than the values of $\alpha$ required by the DIM. Running the simulations clearly shows that, regardless of the initial transport, there is decays as the disc relaxes from its initial conditions. 

\end{enumerate}

We conclude that hydrodynamic spiral waves appear unable to explain accretion in quiescent dwarf novæ and low mass X-ray binaries (whose outer discs sample similar densities and temperatures).

\begin{acknowledgements}
     
The authors would like to thank the anonymous referee for their comments that helped them improve this work. GL acknowledges support from the European Research Council (ERC) under the European Union Horizon 2020 research and innovation program (Grant agreement No. 815559 (MHDiscs)). The simulations and data reduction used in this article were performed using the GRICAD infrastructure (\url{https://gricad.univ-grenoble-alpes.fr}), which is supported by the Grenoble research communities, as well as the Jean-Zay supercomputer at IDRIS \url{http://www.idris.fr/jean-zay/}. This work was granted access to the HPC resources of IDRIS under the allocation 2021-A0120402231 made by GENCI. Data reduction for this paper was performed by making extensive use of the SciPy \citep{virtanen_scipy_2020}, NumPy \citep{harris_array_2020}, pandas \citep{reback_pandas-devpandas_2022} and Matplotlib \citep{hunter_matplotlib_2007} Python libraries. \idefix\ uses the \textsc{Kokkos} portability tool \citep{carter_edwards_kokkos_2014}.

\end{acknowledgements}

%
%
\bibliography{formatedbib}
\bibliographystyle{aa}

\begin{appendix} 
\section{Linear theory wavelength}

\begin{figure}[h]
\centering
\includegraphics[width=\hsize]{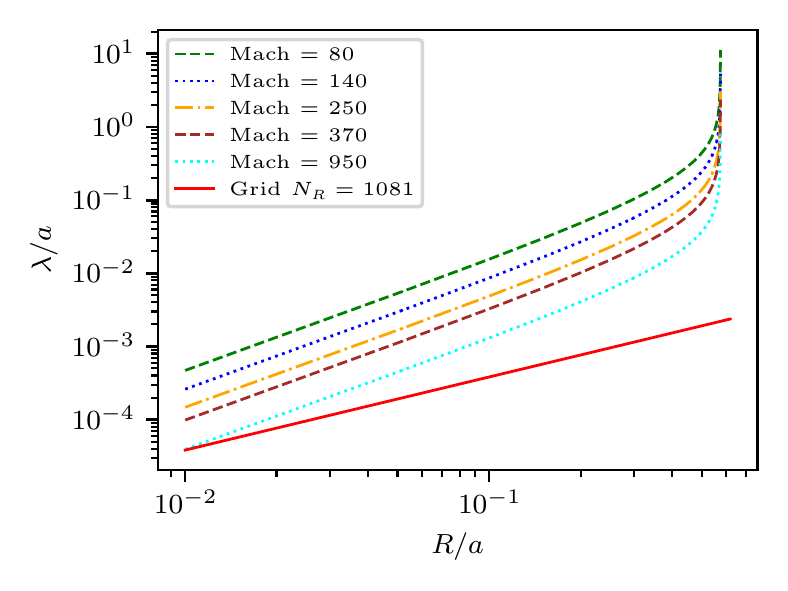}
  \caption{Grid spacing versus analytical wavelength for the spiral mode $m=2$ as a function of the radius.}
  \label{fig:resolution}
\end{figure}

The wavelength from linear theory is shortest at the inner edge of the disc (Eq. \ref{eq:lambda}). Figure \ref{fig:resolution} shows how this wavelength compares with the grid spacing for our `low-resolution' runs. Our resolution (with 1081 radial points) is sufficiently fine up to much higher Mach than explored in this work. In practice, the $m=1$ mode that develops has a much wider wavelength. This plot also shows that the width of the wave-killing zones implemented at the inner boundary is much wider than the linear theory $m=2$ wavelength by several orders of magnitude (see Sect. \ref{sec:spiral}).

\FloatBarrier

\section{Reproducibility with \idefix}
\label{app:ju}
\begin{figure}[h]
\centering
\includegraphics[width=\hsize]{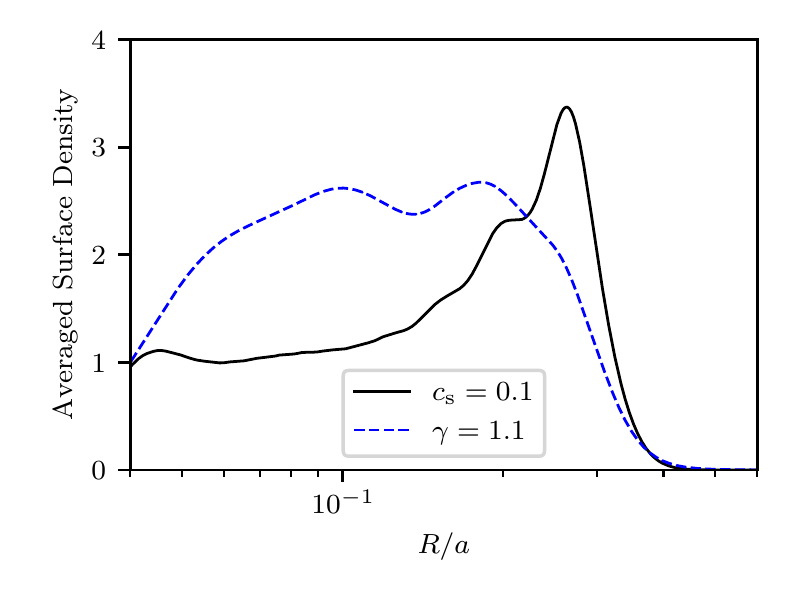}
  \caption{Density profile after four binary orbits, averaged over four binary orbits ($t = 25$ to $t=50$ using the units of \citealt{ju_global_2016}).}
  \label{fig:judensity}
\end{figure}

\begin{figure}[h]
\centering
\includegraphics[width=\hsize]{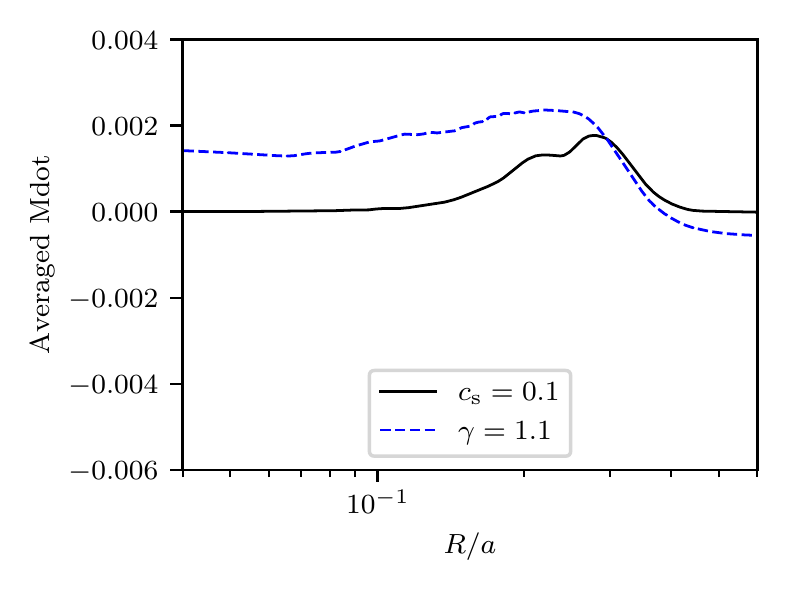}
  \caption{Same as figure \ref{fig:judensity} but for the accretion rate.}
  \label{fig:jumdot}
\end{figure}

\begin{figure}[h]
\centering
\includegraphics[width=\hsize]{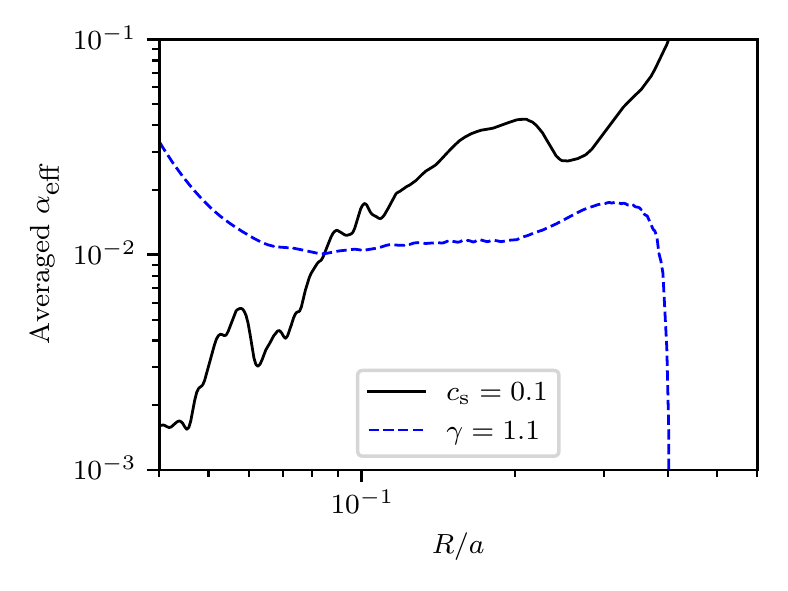}
  \caption{Same as figure \ref{fig:judensity} but for the effective $\alpha$.}
  \label{fig:jualpha}
\end{figure}

We verified that we could reproduce the results of \cite{ju_global_2016} with the code \idefix. Figures \ref{fig:judensity}, \ref{fig:jumdot}, and \ref{fig:jualpha} can be compared with their Figure 12. Although we only present isothermal simulations in the main text, we also compared non-isothermal runs with an adiabatic index of $\gamma=1.1$.  The general agreement is very good, with small differences likely due to variations in implementation between \idefix\ and \athena. 

As the simulations presented in this paper do not use the same time units (by a factor of $2\pi$), the units in this section are converted to the same units as in \cite{ju_global_2016}. We also note that: (1) the $\alpha$ parameter plotted on Fig. \ref{fig:jualpha} is computed in the same way as in \cite{ju_global_2016}, \textit{i.e.} from the radial mass flux; (2) our simulation box extends down to $R = 0.02$ but the plot only extends down to $R=0.04$    to  ease comparison with \cite{ju_global_2016}; and (3) the Mach numbers cannot be directly compared because of differences in the location of the inner boundary (where it is defined).

\FloatBarrier

\section{Definition(s) of $\alpha$}
\label{app:alpha}

\begin{figure}[h]
\centering
\includegraphics[width=\hsize]{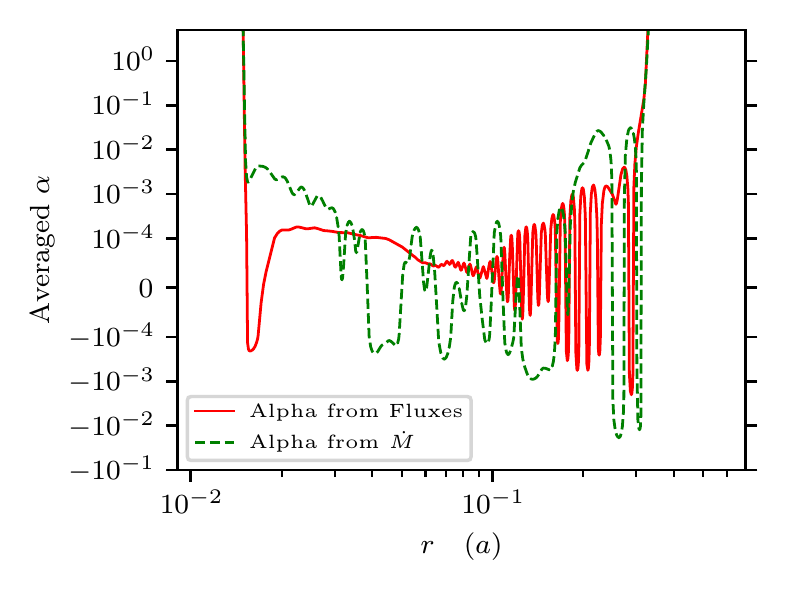}
  \caption{Angular momentum transport parameter $\alpha$ averaged over azimuth and ten binary orbits for the run $\mathrm{Ma} = 250$ and $q=0.3$. The definition from fluxes (this paper) is plotted  with a solid red line and the definition from radial mass flux \citep{ju_global_2016} is plotted as a green dashed line.}
  \label{fig:alphadef}
\end{figure}

The angular momentum transport parameter can be defined in several ways, and not all definitions are equivalent, depending on the exact system. \cite{ju_global_2016} argue that for binary systems such as the one studied here, this parameter should be defined from the accretion rate rather than from the Reynolds (and, in MHD, Maxwell) stress. It has been claimed that this effective transport parameter $\alpha_\text{eff}$  better reflects the influence of the tidal torque of the companion star. \cite{ju_global_2016} take
\begin{equation}
    \alpha_\text{eff} = \frac{\dot{M}}{3\pi\Sigma c_\text{s} H}
,\end{equation}
where the accretion rate $\dot{M}$ is taken from the inter-cell fluxes computed by the Riemann solver at each time step. We chose to measure $\alpha$ from the angular momentum flux (Eq. \ref{eq:alpha}). Both definitions can be obtained from the angular momentum balance equation, using several approximations. In the case of the definition from the radial mass flux, there is a clear contribution from the tidal torque. In our simulations, the disc is clearly not in steady state, and so a definition of the transport parameter based on the accretion rate $\dot{M}$ cannot easily be decomposed into a Reynolds stress term (e.g. our $\alpha$) plus a tidal torque term, as this would require $\dot{M}$ to be constant with respect to the radius. In practice, there is indeed a slight difference in the values computed using these two methods; however, these differences do not change the conclusions of this paper. As shown in Figure \ref{fig:alphadef}, regardless of the chosen definition, we obtain $\alpha \ll 10^{-2}$ at late times.

\section{Pitch angle measurements}
\label{app:pitch}
The pitch angle $\theta$ of a spiral $R(\phi)$ is defined as $\tan \theta = \frac{R'}{R}$. From the Fourier decomposition of Eq. (\ref{eq:TF}), we can define the spiral associated to each mode $m$ as $\phi_m(R) = \arg (\tilde{\Sigma}_m)$ (up to a constant phase). Therefore, 
\begin{equation}
    \theta_m = \arctan \left ( \frac{1}{R \phi_m'(R)} \right )
.\end{equation}

\begin{figure*}[h]
\centering
\includegraphics[width=\hsize]{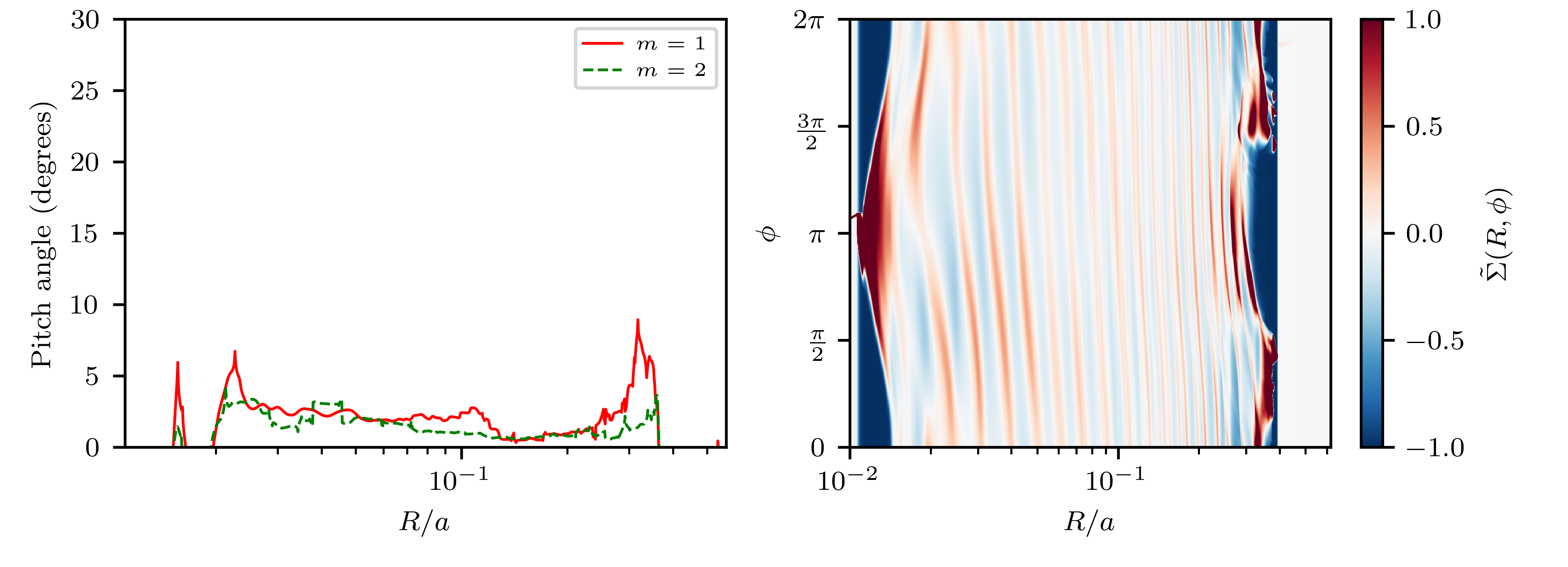}
  \caption{Density profile properties of the run $\textrm{Ma} = 250$ after 38.8 binary orbits. Left panel: Pitch angle of the $m=1$ and $m=2$ spiral modes in the disc in the density profile. Right panel: De-projected density fluctuation $\tilde{\Sigma}$, as defined in Eq. (\ref{eq:TF}).}
  \label{fig:pitch}
\end{figure*}

Figure \ref{fig:pitch} shows the spiral pattern on a deprojected plot. Consistently with Fig. \ref{fig:DensityTF}, we see that the one-armed spiral wave dominates the dynamic. The left panel shows that pitch angle of the spiral wave is consistently $< 5^\circ$ in the disc. The spiral waves are tightly wound. This explains how, even though the $m=1$ mode dominates, the disc remains circular in shape
overall, rather than very eccentric.

\FloatBarrier
\section{RWI/KHI-driven accretion}

\begin{figure*}[h]
\centering
\includegraphics[width=\hsize]{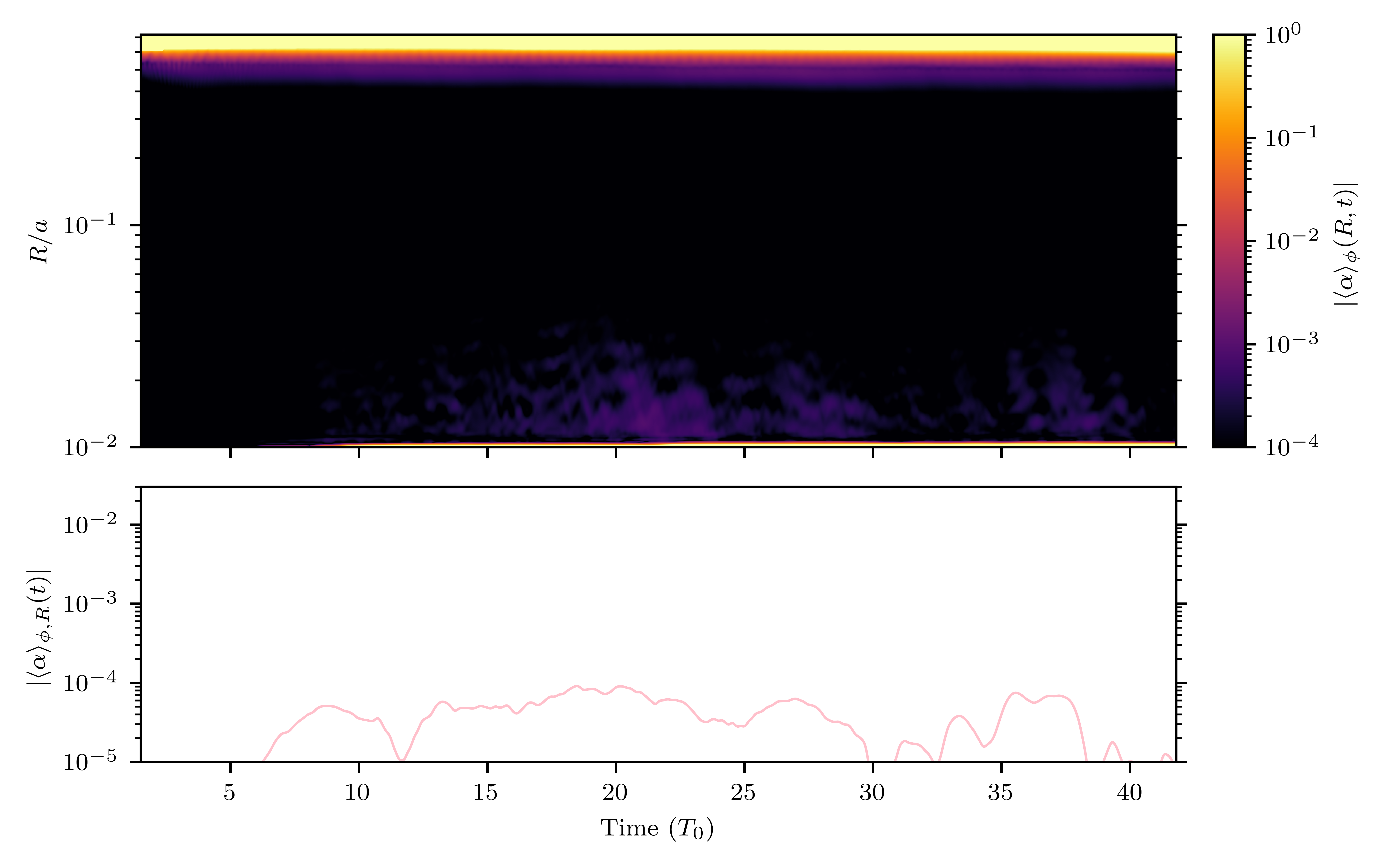}
  \caption{Measured angular momentum transport parameter.  Top panel: Measured angular momentum transport parameter $\alpha$ azimuthally averaged and smoothed over three binary orbits for a run with $\mathrm{Ma} = 250$ and no companion star. Here we plot the absolute value of the transport parameter. Bottom panel: Azimuthally and radially (over $0.02 \leq R \leq 0.1$) averaged $\alpha$, smoothed over three binary orbits for three runs. Compared to Fig. \ref{fig:KP2D}, the $\alpha$ scale of the bottom panel has been extended to show lower values of the transport parameter.}
     \label{fig:NoSalpha}
\end{figure*}

As discussed above, we verified that the mixture of RWI/KHI and centrifugal instability drives negligible accretion compared to the spiral shocks triggered by the tidal potential of the secondary star. The setup used to produce Fig. \ref{fig:NoSalpha} is the same as for the run with $\mathrm{Ma} = 250$ and $q=0.1$. The mass of the white dwarf is the same but the secondary mass $M_\textrm{s}$ is set to zero in the potential (Eq. \ref{eq:potential}). This setup is very similar to the setup of \cite{lesur_spiral-driven_2015}.

\end{appendix}

\end{document}